\documentclass[12pt,preprint]{aastex}

\begin{document}

\title{The active W UMa type binary star V781 Tau revisited}

\author{Li K.\altaffilmark{1,2}, Gao, D.-Y.\altaffilmark{1}, Hu, S.-M.\altaffilmark{1}, Guo, D.-F.\altaffilmark{1}, Jiang, Y.-G.\altaffilmark{1}, Chen, X.\altaffilmark{1}}

\altaffiltext{1}{Shandong Provincial Key Laboratory of Optical Astronomy and Solar-Terrestrial Environment, Institute of Space Sciences, Shandong University,Weihai, 264209, China (e-mail: gaodongyang@sdu.edu.cn (Gao, D.-Y.), husm@sdu.edu.cn (Hu, S.-M.))}
\altaffiltext{2}{Key Laboratory for the Structure and Evolution of Celestial Objects, Chinese Academy of Sciences}

\begin{abstract}
In this paper, new determined $BVR_cI_c$ light curves and radial velocities of V781 Tau are presented. By analyzing the light curves and radial velocities simultaneously, we found that V781 Tau is a W-subtype medium contact binary star with a mass ratio of $q=2.207\pm0.005$ and a contact degree of $f=21.6(\pm1.0)\%$. The difference between the two light maxima was explained by a dark spot on the less massive primary component. The orbital period change of V781 Tau was also investigated. A secular decrease at a rate of $-6.01(\pm2.28)\times10^{-8}$ d/yr and a cyclic modulation with a period of 44.8$\pm$5.7 yr and an amplitude of $0.0064\pm0.0011$ day were discovered. The continuous period decrease may be caused by angular momentum loss due to magnetic stellar wind. Applegate mechanism failed to explain the cyclic modulation. It is highly possible that the cyclic oscillation is the result of the light travel time effect by a third companion.

\end{abstract}


\section{Introduction}
The variability of V781 Tau was first discovered by Harris (1979) according to his two nights photoelectric observations. The light curve of V781 Tau shows typical W UMa type. Two values of orbital period, 0.$^d33939$ and 0.$^d 34494$, were determined by Harris (1979), but he can not point out the more accurate one. Later, Berthold (1981, 1983) analyzed his photographic observations and gave an ephemeris: Min. (HJD) = 2443874.954 + 0.$^d 3449100 \times$ E. Diethelm (1981) also preferred the longer period by analyzing his photoelectric measurements. After that, several authors published many times of minimum light. Cereda et al. (1988) carried out extensive photoelectric observations in B and V bands. With the Fourier analysis technique proposed by Niarchos (1983), Cereda et al. (1988) obtained the inclination of the orbit to be $68^\circ\pm2^\circ$. The first radial velocity observations of V781 Tau were started by Lu (1993), who reanalyzed the BV light curves of Cereda et al. (1988) combining his radial velocities, the absolute dimensions were determined. Yang \& Liu (2000) investigated the orbital period variation of V781 Tau and found that the period of V781 tau is secular decrease with a rate of $dP/P=5.0\times10^{-11}$. New radial velocities of V781 Tau were obtained by Zwitter et al. (2003). They simultaneously analyzed the Hipparcos and Tycho photometry and radial velocity data. Recently, Yakut et al. (2005) and Kallrath et al. (2006) observed and analyzed V781 Tau. They all obtained the basic physical parameters and the orbital period change behavior.

It has been nine years since the last investigation of V781 Tau, we started photometric and spectroscopic observations in order to determine the physical parameters and the orbital period variation.

\section{Photometric and Spectroscopic Observations}

Photometric observations of V781 Tau were carried out using the 1.0-m telescope at Weihai Observatory of Shandong University (Hu et al. 2014) on November 4, 2013 and December 24, 2014. The observation made in 2013 was using a back-illuminated PIXIS 2048B CCD camera attached to the Cassegrain telescope, while an Andor DZ936 CCD camera was used in 2014. Both of the two CCD cameras have $2048\times2048$ square pixels (13.5$\times$13.5$\mu$m pixel$^{-1}$), resulting an effective CCD field of about 11.8$'$ $\times$ 11.8$'$. The filter system is a standard Johnson-Cousin-Bessel $UBVR_cI_c$ CCD photometric system. Each of the CCD images was differently exposed from 2 s to 5 s based on the used filters. During the observations, GSC 01870-00514 ($\alpha_{2000.0}=05^h50^m22^s.39$, $\delta_{2000.0}=+26^{\circ} 59^{\prime}55.0^{\prime\prime}$, $V=9.68$, $B-V=0.45$ ) and 2MASS J05502287+2654060 ($\alpha_{2000.0}=05^h50^m22^s.87$, $\delta_{2000.0}=+26{\circ} 54^{\prime}06.1^{\prime\prime}$, $V=10.96$, $B-V=0.95$ ) were chosen as the comparison and check stars, respectively. All the measured images were firstly corrected with bias and flat images and then processed using the aperture photometry (APPHOT) package in the Image Reduction and Analysis
Facility (IRAF\footnote{IRAF is distributed by the National Optical Astronomy Observatories, which are operated by the Association of Universities for Research in Astronomy under cooperative agreement with the National Science Foundation.}). The observations on November 4, 2013 determined a time of minimum light only. Complete light curves were determined on December 24, 2014, the corresponding phased light curves are displayed in Figure 1 and the original photometric data are shown in Table 1. The phases are calculated using the following ephemeris:
\begin{eqnarray}
Min.I = HJD2457016.32177 + 0.^d 34490986E,
\end{eqnarray}
the period in this equation is taken from Kreiner (2004).
Three times of minimum light are determined during the two observing nights, they are: 2456601.3948$\pm$0.0002, 2457016.1504$\pm$\\0.0002 and 2457016.3208$\pm$0.0003.

B and V light curves of V781 Tau derived by Cereda et al. (1988), Yakut et al. (2005), Kallrath et al. (2006) and us are shown in Figure 2. In order to derive the change of the light curves, we shifted other light curves to our observations at the primary minimum. As seen in Figure 2, the light curve of V781 Tau is changed significantly and shows positive type of O'Connell effect (O'Connell 1951). The differences between the two light maximum of the four sets of light curves exhibit continuously variation. This may indicate that V781 Tau shows very strong magnetic activity.

We used the Weihai Echelle Spectrograph (WES) to obtain spectra of V781 Tau for  radial velocity  calculation. We obtained seven spectra of V781 Tau and a spectra of HD50692 as Radial Velocity Standard Star with spectral type of G0V on February 4, 2015.
The raw data were reduced using the echelle package. Radial velocities of V781 Tau were obtained using task fxcor in IRAF, with the spectrum of HD50692 as the template. The spectra from WES  has 107 orders because of echelle spectrograph configuration. We calculated the arithmetic mean value of radial velocities from different spectral orders, and set the standard deviation divided by the square root of numbers as configuration error. The determined radial velocities of V781 Tau are listed in Table 2.

\begin{table*}
\begin{center}
\caption{Original photometric data of V781 Tau observed on December 24, 2014, Hel. JD 2457000+}
\begin{tabular}{cccccccc}\hline\hline
Hel. JD&$\Delta m$&Hel. JD&$\Delta m$&Hel. JD&$\Delta m$&Hel. JD&$\Delta m$\\\hline
\multicolumn{2}{c}{B}&\multicolumn{2}{c}{V}&\multicolumn{2}{c}{R$_c$}&\multicolumn{2}{c}{I$_c$}\\\hline
15.9821& 	-0.442&   15.9824& 	-0.560&   15.9826& 	-0.633&   -0.7140& 	-0.367\\
15.9830& 	-0.451&   15.9833& 	-0.575&   15.9834& 	-0.632&   -0.7310& 	-0.359\\
15.9839& 	-0.449&   15.9842& 	-0.564&   15.9843& 	-0.654&   -0.7230& 	-0.375\\
15.9848& 	-0.470&   15.9850& 	-0.573&   15.9852& 	-0.661&   -0.7210& 	-0.363\\
15.9857& 	-0.467&   15.9859& 	-0.586&   15.9861& 	-0.656&   -0.7420& 	-0.360\\
15.9865& 	-0.476&   15.9868& 	-0.580&   15.9869& 	-0.681&   -0.7390& 	-0.371\\
15.9874& 	-0.476&   15.9877& 	-0.588&   15.9879& 	-0.677&   -0.7490& 	-0.366\\
15.9883& 	-0.478&   15.9886& 	-0.604&   15.9887& 	-0.672&   -0.7450& 	-0.362\\
\hline

\end{tabular}
\end{center}
(The full version of this table is available in electronic form in the online journal. A portion is shown here for guidance regarding its form and content.)
\end{table*}

\begin{figure}
\begin{center}
\includegraphics[angle=0,scale=1]{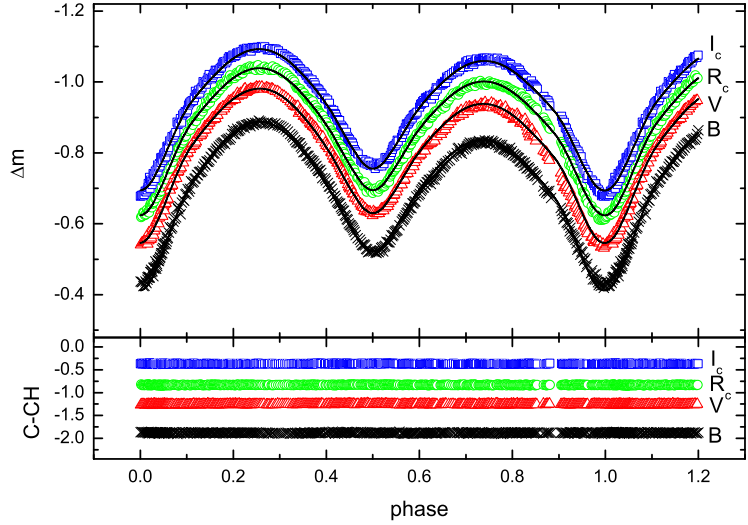}
\caption{The observed four color light curves of V781 Tau. Different colors represent different filters.}
\end{center}
\end{figure}

\begin{figure*}
\begin{center}
\includegraphics[angle=0,scale=0.9]{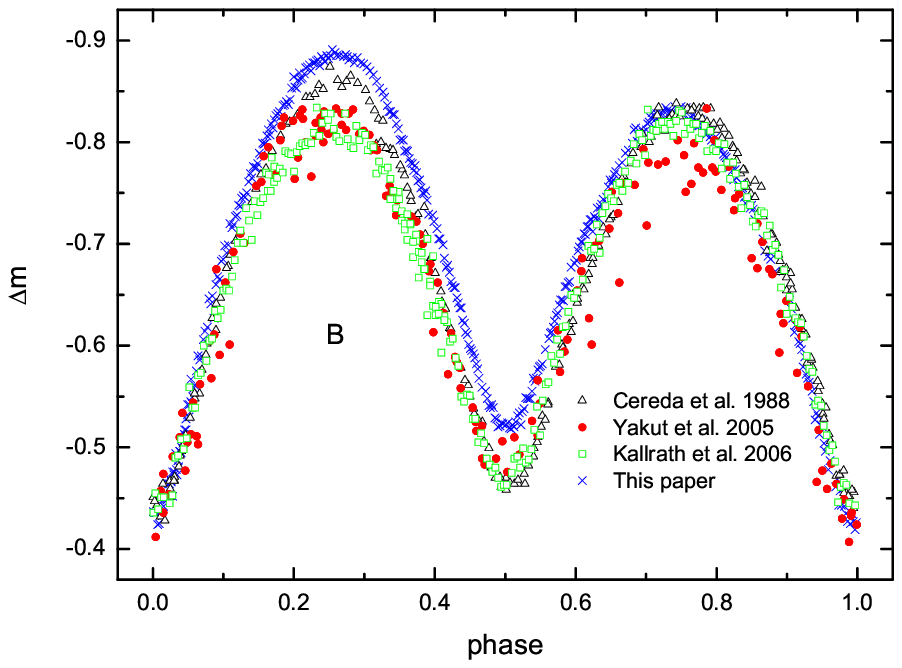}\\
\includegraphics[angle=0,scale=0.9]{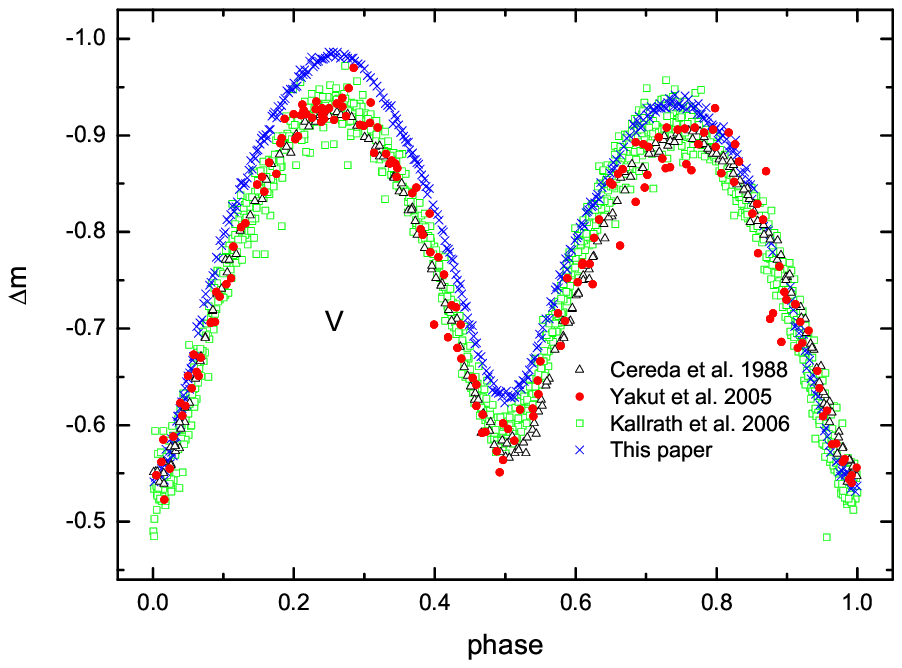}

\caption{Top panel shows the light curves comparison in B band, while the lower panel displays V band.}
\end{center}
\end{figure*}

\begin{table}
\begin{center}
\caption{Heliocentric Radial Velocities of V781 Tau}
\begin{tabular}{lccccc}
\hline\hline
 JD (Hel.) &  Phase$^*$  &RV$_1$&  Errors  & RV$_2$ & Errors  \\
2457000+   &         & km s$^{-1}$& km s$^{-1}$ & km s$^{-1}$ &km s$^{-1}$ \\\hline
57.98872   & 0.8053  &  258.2    &  $\pm$17.8    & -71.9  &  $\pm$8.7   \\
58.02855   & 0.9208  &  147.6    &  $\pm$16.1    & -41.1  &  $\pm$15.0  \\
58.05684   & 0.0028  &  26.3     &  $\pm$6.9     & 26.3   &  $\pm$6.9   \\
58.08512   & 0.0848  &  -123.2   &  $\pm$18.1    & 93.5   &  $\pm$9.8   \\
58.14682   & 0.2637  &  -215.6   &  $\pm$16.6    & 141.8  &  $\pm$7.2   \\
58.17511   & 0.3457  &  -170.9   &  $\pm$8.9     & 118.6  &  $\pm$9.0   \\
58.20350   & 0.4280  &  -84.5    &  $\pm$21.8    & 85.4   &  $\pm$25.3  \\
\hline
\end{tabular}
\end{center}
\scriptsize
$^*$ Phases are computed with the Equation (1).
 \end{table}

\section{Photometric solutions}

We used the W-D program (Wilson \& Devinney 1971; Wilson 1990, 1994) to analyze the four color light curves and radial velocities of V781 Tau simultaneously.
The $(B-V)_0\approx0^m.55$ was determined by Kallrath et al. (2006), which is corresponding to a spectral type of G0V. Therefore, the effective temperature of the primary component of V781 Tau was set to be $T_1=6000$ K. Accordingly, the gravity-darkening coefficients and bolometric albedo coefficients of the two components were fixed at $g_{1,2}=0.32$ $A_{1,2}=0.5$ based on Lucy (1967) and Ruci\'{n}ski (1969), respectively. The limb darkening coefficients were adopted from the limb darkening table of Van Hamme (1993). Iterative studies have shown that V781 Tau is a contact binary. Mode 3 was used during the solutions. The orbital inclination $i$, the mass ratio $q$, the effective temperature of the second component $T_2$, the monochromatic luminosity in each band of the primary component $L_1$ and the dimensionless potentials of the two components, $\Omega_1 = \Omega_2$ are adjustable parameters.

As shown in Figure 2, the light curve of V781 Tau is seasonally changed, the spot mode of the W-D programme was used. Extensive investigations reveal that a dark spot on the less massive primary component leads to the best fit. The solution results are listed in Table 3. The spectroscopic orbital elements are shown in Table 4. The radial velocities and the comparison between observed and synthetic light curves are displayed in Figures 3 and 1, respectively. The corresponding geometric structure at phase 0.75 is plotted in Figure 4. As seen in Figure 1, the theoretical light curves show a mismatch in the primary minimum. We tried to reanalyze the light curves with two or more spots. However, very similar results are obtained. Therefore, we investigated the $B$, $V$, $R_c$ and $I_c$ light curves individually. The derived photometric elements are listed in Table 3, the corresponding fitted light curves are shown in Figure 5, in which very well fit at the primary minimum can be seen.

\begin{figure}
\begin{center}
\includegraphics[angle=0,scale=1.0]{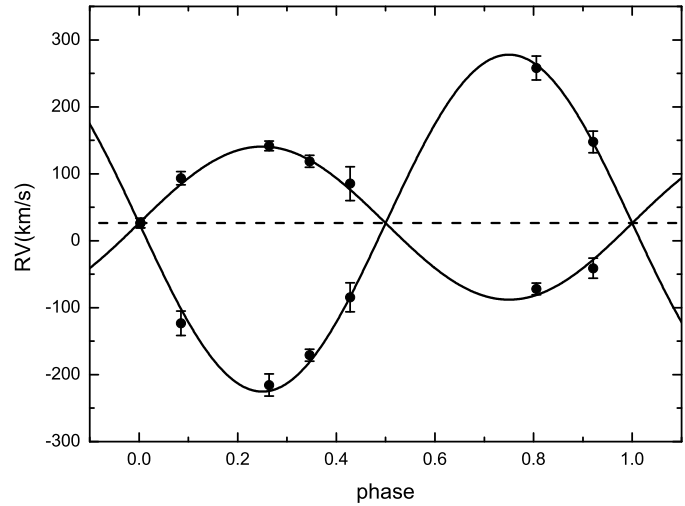}
\caption{ The phased radial velocity curves for V781 Tau. The solid line shows a sine fit.}
\end{center}
\end{figure}

\begin{figure}
\begin{center}
\includegraphics[angle=0,scale=1.0]{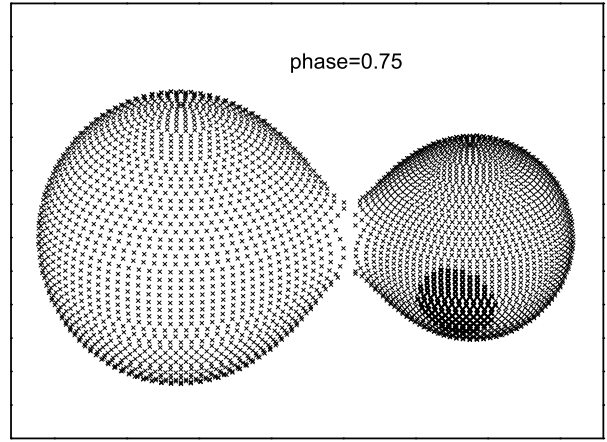}
\caption{Geometrical configuration at phase 0.75. }
\end{center}
\end{figure}
\begin{figure}
\begin{center}
\includegraphics[angle=0,scale=1.0]{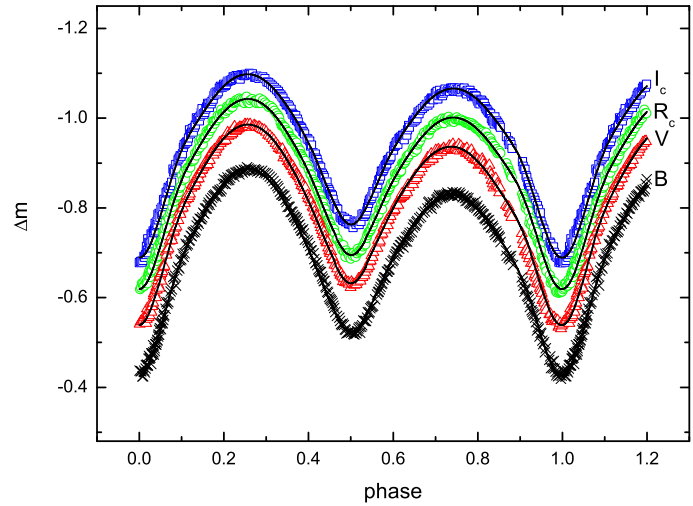}
\caption{Observed and synthetic light curves derived by analyzing the four color light curves individually. The synthetic light curves fit very well at the primary minimum. }
\end{center}
\end{figure}

\begin{table}
\small
\begin{center}
\caption{Photometric solutions for V781 Tau}
\begin{tabular}{lccccc}
\hline
Parameters &  BVR$_c$I$_c$ & B & V & R$_c$ & I$_c$  \\

\hline
    $ g_1=g_2$ &\multicolumn{5}{c}{0.32$^a$} \\
     $A_1=A_2$ &\multicolumn{5}{c}{0.5$^a$} \\
     $x_{1bol},$ $x_{2bol}$&  \multicolumn{5}{c}{0.647, 0.649$^a$}\\

     $y_{1bol},$ $y_{2bol}$& \multicolumn{5}{c}{0.221, 0.193$^a$}\\

     $x_{1B},$ $x_{2B}$& \multicolumn{5}{c}{0.829, 0.847$^a$}\\

     $y_{1B},$ $y_{2B}$& \multicolumn{5}{c}{0.185, 0.098$^a$}\\

     $x_{1V},$ $x_{2V}$& \multicolumn{5}{c}{0.745, 0.778$^a$}\\

     $y_{1V},$ $y_{2V}$& \multicolumn{5}{c}{0.256, 0.200$^a$}\\

     $x_{1R_c},$ $x_{2R_c}$& \multicolumn{5}{c}{0.674, 0.708$^a$}\\

     $y_{1R_c},$ $y_{2R_c}$& \multicolumn{5}{c}{0.269, 0.229$^a$}\\

     $x_{1I_c},$ $x_{2I_c}$& \multicolumn{5}{c}{0.590, 0.623$^a$}\\

     $y_{1I_c},$ $y_{2I_c}$& \multicolumn{5}{c}{0.260, 0.230$^a$}\\

    $ T_1(K) $ &  \multicolumn{5}{c}{6000$^a$}   \\

  $q(M_2/M_1) $&  \multicolumn{5}{c}{2.207$ \pm0.005$}    \\

  $ \Omega_{in} $ & \multicolumn{5}{c}{5.5420$^a$}  \\

  $ \Omega_{out} $ &\multicolumn{5}{c}{4.9389$^a$} \\

        $T_2(K) $&     5575$ \pm5$ & $5600\pm7$ &   $5525\pm10$ &     $5537\pm9$ &    $5494\pm10$  \\

         $i$ &  $65.9\pm0.2$ & $66.4\pm0.2$ & $66.6\pm0.2$ & $66.7\pm0.2$& $66.4\pm0.2$  \\

$L_{1B}/L_B$&   0.4421$ \pm0.0008$ & $0.4361\pm0.0010$ &   &   &    \\
$L_{1V}/L_V$&   0.4128$ \pm0.0006$ &   & $0.4240\pm0.0012$ &   &    \\
$L_{1R_c}/L_{R_c}$&  0.3990$ \pm0.0005$ &   &   & $0.4058\pm0.0009$ &    \\
$L_{1I_c}/L_{I_c}$&  0.3889$ \pm0.0005$ &   &  &   & $0.4001\pm0.0007$  \\

  $\Omega_1$=$\Omega_2$ & $5.3956\pm0.0039$ & $5.3897\pm0.0062$ & $5.4041\pm0.0041$ & $5.4255\pm0.0039$  \\

  $r_1(pole)$ &   0.3031$ \pm 0.0006$ &$0.3045\pm0.0003$ & $0.3050\pm0.0005$ & $0.3038\pm0.0004$ & $0.3019\pm0.0003$  \\

  $r_1(side)$ &   0.3179$ \pm0.0007$ & $0.3196\pm0.0004$ & $0.3202\pm0.0007$ & $0.3187\pm0.0004$ & $0.3164\pm0.0004$  \\

  $r_1(back)$ &   0.3585$ \pm0.0013$ & $0.3614\pm0.0007$ & $0.3625\pm0.0011$ & $0.3599\pm0.0007$ & $0.3561\pm0.0007$  \\

  $r_2(pole)$ &   0.4330$ \pm0.0005$ & $0.4343\pm0.0003$ & $0.4248\pm0.0005$ & $0.4336\pm0.0003$ & $0.4319\pm0.0003$  \\

  $r_2(side)$ &   0.4634$\pm0.0007$ &  $0.4765\pm0.0004$ & $0.4658\pm0.0007$ & $0.4642\pm0.0005$ & $0.4619\pm0.0004$  \\

  $r_2(back)$ &    0.4951$ \pm0.0009$ &$0.4975\pm0.0010$ & $0.4983\pm0.0009$ & $0.4962\pm0.0006$ & $0.4931\pm0.0006$  \\

  $f$& 21.6$\pm1.0\%$&$24.3\pm0.6\%$ &  $25.3\pm1.0\%$ &  $22.9\pm0.7\%$ & $19.3\pm0.6\%$  \\

  $\theta(radian)$ &  1.90$ \pm0.19$& $1.87\pm0.16$ & $1.90\pm0.16$ & $1.90\pm0.16$ & $1.88\pm0.16$  \\

  $\phi(radian)$ & 1.40$ \pm0.15$  &$1.48\pm0.11$ & $1.42\pm0.11$ & $1.57\pm0.11$ & $1.56\pm0.11$  \\

  $r(radian)$ & 0.44$ \pm0.06$     &$0.44\pm0.04$ & $0.44\pm0.04$ & $0.44\pm0.04$ & $0.41\pm0.04$  \\

  $T_f(T_d/T_0)$ &  0.75$ \pm0.07$ &$0.69\pm0.07$ & $0.71\pm0.08$ & $0.74\pm0.08$ & $0.75\pm0.07$  \\
\hline
\end{tabular}
\end{center}
$^a$ Assumed parameters
\end{table}

\begin{table}
\begin{center}
\caption{The spectroscopic orbital elements for V781 Tau}
\begin{tabular}{lcl}
\hline
Parameters &  Values & Errors  \\

\hline

  $V_0(km/s)$ &     26.3&    $\pm0.5$\\

  $K_1(km/s)$ &      251.6&    $\pm15.9$\\

  $K_2(km/s)$ &      114.0&     $\pm13.8$\\

  $a\sin i(R_\odot)$ &      2.48&    $\pm0.16$\\

  $M_1\sin ^3i(M_\odot)$ &      0.54&    $\pm0.05$\\

  $M_2\sin ^3i(M_\odot)$ &      1.19&     $\pm0.08$\\

\hline
\end{tabular}
\end{center}
\end{table}

\section {Period investigation}	
The latest analysis of the orbital period variation of V781 Tau was started by Kallrath et al. (2006). It has been more than nine years. Many times of minimum light has been determined during this period. So, we collected all available times of minimum light to analyze the orbital period changes. A total of 220 timings of minimum light, including ours, were compiled from literatures, and they are listed in Table 5. The $(O-C)_1$ values were calculated using the linear ephemeris determined by Kreiner (2004),
 \begin{equation}
\textrm{Min.I}=2452500.0720+0.^d 34490986\textrm{E}.
\end{equation}

The $(O-C)_1$ values computed with Equation (2) are listed in the fourth column of Table 5 and plotted in the top panel of Figure 6. We found that the $(O-C)_1$ curve contains a continuous period decrease and a cyclic change. We used the following equation given by Irwin (1952)
\begin{eqnarray}
(O-C)_1=T_0+\Delta T_0+(P_0+\Delta P_0)E+{\beta \over 2}E^2+A[(1-e^2){\sin(\nu+\omega)\over(1+e\cos\nu)}+e\sin\omega] \nonumber\\
=T_0+\Delta T_0+(P_0+\Delta P_0)E+{\beta \over 2}E^2+A[\sqrt{(1-e^2 )}\sin E^*\cos\omega+\cos E^*\sin \omega],
\end{eqnarray}
to fit the $(O-C)_1$ values. In this equation, $T_0$ and $P_0$ are, respectively, the initial epoch and the orbital period, $\Delta T_0$ and $\Delta P_0$ are their corrections, and $\beta$ is the long-term period change. Other parameters were taken from Irwin (1952), which can be determined by the Levenberg-Marquart method (Press et al. 1992). In the calculation, the weights for the visual and photographic minima were 1 and that for the photoelectric and CCD minima were 8. The final solution parameters are listed in Table 6. A $44.8\pm5.7$ yr cyclic variation superimposed on a long-term period decrease at a rate of $-6.01(\pm2.28)\times10^{-8}$ d/yr was discovered. When the long-term period decrease was removed, the $(O-C)_2$ values are displayed in the middle panel of Figure 6. After the full ephemeris were subtracted, the residuals are plotted in the lowest panel of Figure 6.

\begin{table*}
\begin{center}
\caption{Times of minimum light for V781 Tau}
\begin{tabular}{lccrcccr}\hline\hline
JD Hel. & Method & Type &  E & $(O-C)_1$ & $(O-C)_2$ &Residuals &  Reference \\\hline
32881.4600  & pg  & s & -56880.5 & 0.0333   & -0.0022  & -0.0041  & IBVS 2443              \\
33950.5150  & pg  & p & -53781   & 0.0402   & 0.0035   & 0.0036   & IBVS 2443              \\
34775.3680  & pg  & s & -51389.5 & 0.0413   & 0.0039   & 0.0058   & IBVS 2443              \\
35540.3710  & pg  & s & -49171.5 & 0.0342   & -0.0037  & -0.0001  & IBVS 2443              \\
36610.2850  & pg  & s & -46069.5 & 0.0378   & -0.0005  & 0.0052   & IBVS 2443              \\
36637.3350  & pg  & p & -45991   & 0.0124   & -0.0259  & -0.0201  & MHAR 16.10             \\
36957.4420  & pg  & p & -45063   & 0.0430   & 0.0046   & 0.0108   & IBVS 2443              \\
38002.4790  & pg  & p & -42033   & 0.0031   & -0.0354  & -0.0307  & MHAR 16.10             \\
38088.3970  & pg  & p & -41784   & 0.0386   & 0.0001   & 0.0045   & IBVS 2443              \\
38440.3780  & pg  & s & -40763.5 & 0.0391   & 0.0006   & 0.0038   & IBVS 2443              \\
39536.3270  & pg  & p & -37586   & 0.0370   & -0.0012  & -0.0016  & IBVS 2443              \\
40981.3320  & pg  & s & -33396.5 & 0.0421   & 0.0047   & 0.0011   & IBVS 2443              \\
41329.3450  & pg  & s & -32387.5 & 0.0411   & 0.0040   & -0.0002  & IBVS 2443              \\
41330.3810  & pg  & s & -32384.5 & 0.0424   & 0.0053   & 0.0011   & IBVS 2443              \\
41333.3230  & pg  & p & -32376   & 0.0526   & 0.0155   & 0.0113   & MHAR 16.10             \\
41337.2790  & pg  & s & -32364.5 & 0.0422   & 0.0051   & 0.0009   & IBVS 2443              \\
42839.3630  & pg  & s & -28009.5 & 0.0437   & 0.0081   & 0.0023   & IBVS 2443              \\
\hline
\end{tabular}
\end{center}
(This table is available in its entirety in the online journal. A portion is shown here for guidance regarding its form and content.)
 \end{table*}

\begin{figure}
\begin{center}
\includegraphics[angle=0,scale=1.0]{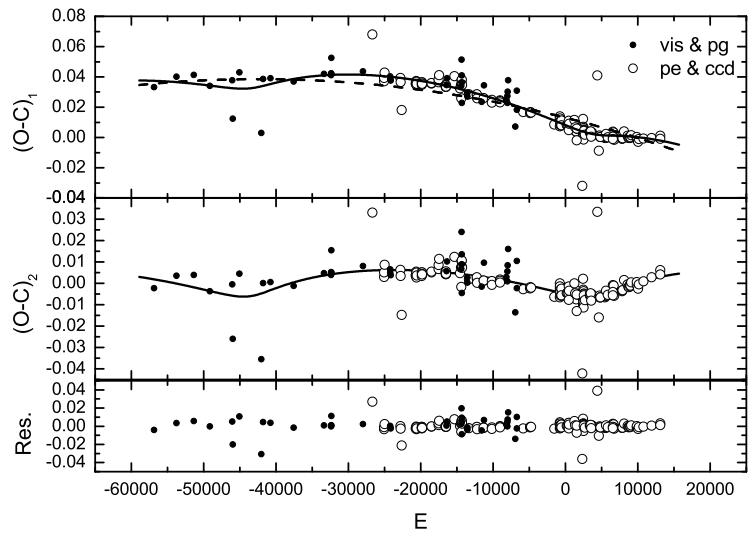}
\caption{$O-C$ diagram of V781 Tau. Top panel shows $(O-C)_1$ curve determined by the Equation (1). The $(O-C)_2$ values which remove the long-term period decrease from the $(O-C)_1$ curve are plotted in the middle. The residuals from full ephemeris of Equation (2) are displayed in the lower panel. Filled circles show visual and photographic minima, while open circles refer to the photoelectric and CCD minima. }
\end{center}
\end{figure}

\begin{table}
\begin{center}
\caption{Parameters for the fit of times of minimum light}
\begin{tabular}{lcccc}
\hline\hline

Parameters &     Values &     Errors &    \\\hline

$\Delta T_0$ (d) &    0.0131 &  $\pm0.0010$     \\

$\Delta P_0$ (d) & $-1.20\times10^{-6}$ & $\pm0.22\times10^{-6}$ \\

$\beta$ (d/yr) & $-6.01\times10^{-8}$ & $\pm2.28\times10^{-8}$  \\

     $A$ (d) &     0.0064 &  $\pm0.0011$  \\

         $e$ &       0.55 &    $\pm0.22$  \\

   $P_3$ (yr) &     44.8 &   $\pm5.7$ \\

$\omega (^\circ)$ &       294.4 &    $ \pm38.6$  \\

$ T_P$ (HJD) &  2437588.2 &   $\pm2510.9 $ \\
\hline\hline
\end{tabular}
\end{center}
\end{table}

\section{Discussion and conclusions}
New CCD light curves and radial velocities of V781 Tau were determined by the 1.0-m telescope at Weihai Observatory of Shandong University. Analyzing the light curves and radial velocities using the W-D code simultaneously, we derived that V781 Tau has a mass ratio of $q=2.207\pm0.005$ and a fill out factor of $f=21.6(\pm1.0)\%$. The system velocity was determined to be $V_0=26.3\pm0.5$ km/s. Combining the solution results and radial velocities of the two components, we can determined the absolute parameters of V781 Tau, they are $a=2.72\pm0.18R_\odot$, $M_1=0.71\pm0.07M_\odot$, $M_2=1.57\pm0.11M_\odot$, $R_1=0.89\pm0.05R_\odot$, $R_2=1.26\pm0.08R_\odot$, $L_1=0.92\pm0.07L_\odot$ and $L_2=1.38\pm0.10L_\odot$.

Using all available times of minimum light, we investigated the orbital period changes of V781 Tau. It is found that the orbital period of V781 Tau is secular decrease at rate of $-6.01(\pm2.28)\times10^{-8}$ d/yr, and has a cyclic modulation with a period of $44.8\pm5.7$ yr. Normally, the secular orbital period decrease is caused by mass transfer from the more massive component to the less massive one or by angular momentum loss via magnetic stellar wind. Assuming that the mass transfer is conservative, we can determined the mass transfer rate using the following equation,
\begin{eqnarray}
{\dot{P}\over P}=-3\dot{M_1}({1\over M_1}-{1\over M_2}).
\end{eqnarray}
A mass transfer rate at $dM_1/dt=7.5(\pm2.8)\times10^{-8}\,M_\odot$ yr$^{-1}$ was derived. The thermal timescale of the more massive component can be computed to be ${GM^2\over RL}\sim4.6\times10^7$ yr, which is about 8 times of the timescale of period decrease $P/(dP/dt)\sim5.7\times10^6$ yr. So, the secular period decrease should be due to the angular momentum loss by magnetic stellar wind.

The cyclic oscillation in the orbital period change of V781 Tau can be caused either by the Applegate mechanism (Applegate 1992) due to magnetically active component(s) or by the light travel time effect due to a third companion.
Using the oscillation period and the amplitude listed in Table 5 and the relation ${\Delta{P}\over P}=-9{\Delta Q\over Ma^2}$ (Lanza \& Rodon\`{o} 2002), we can determine that the variations of the quadrupole moment for the primary component is $\Delta Q_1=9.8\times10^{48}$ g cm$^2$. This value is about two orders of magnitude smaller than typical values $10^{51}$ to $10^{52}$ g cm$^2$ for active close binaries. Applegate model failed to explain the cyclic modulation in the orbital period. Therefore, liking many other contact binaries, e.g., EP And (Lee et al. 2013), MR Com (Qian  et al. 2013) and Li et al. (2014), cyclic period variation of V781 Tau is caused by the light travel time effect due to a tertiary.
Using the mass function,
\begin{equation}
f(m)={(m_3\sin i^\prime)^3\over (m_1+m_2+m_3)^2}={4\pi\over GP^2_3}\times(a_{12}\sin i^\prime)^3,
\end{equation}
we determined $f(m) = 6.79\pm3.5\times 10^{-4}M_\odot$. Then, we can calculated that the smallest mass and the greatest septation of the third companion are $m_3=0.16(\pm0.05)\,M_\odot$ and $18.9(\pm6.5)$ AU, respectively. No third signal was found during the spectral observations and the photometric solutions, so the third body should be a very cool dwarf star or a compact object. The orbital motion of the third body can also lead to a periodic variation of the radial velocity of the central eclipsing pair. Lu (1993) first determined that the system radial velocity is 24.4 km/s, and Zwitter derived a 30.44 km/s value, while our result is 26.3 km/s. It seems that the radial velocity values varied in some way. But only three data points can not support a precise study. More radial velocity observations are needed.

V781 Tau is a W-subtype contact binary. We shows the two components of V781 Tau on the mass-luminosity diagram in Figure 7. 42 W-subtype low-temperature contact binary systems (LTCBs) given by Yakut \& Eggleton (2005) are also plotted in Figure 7, where open triangles and circles represent the primary and the secondary of these systems, respectively. In Figure 7, zero age main sequence (ZAMS) and the terminal age main sequence (TAMS) which are constructed by the binary star evolution code (i.e., BSE Code; Hurley et al. 2002) are displayed with continuous and dotted lines, respectively. As seen in Figure 7, the locations of the two components of V781 Tau is similar with other W-subtype LTCBs. The more massive secondary component is under-luminous, while the less massive primary component is over-luminous and over-sized with the respect to their ZAMS masses. This might be the consequence of the assumption that the more massive component is transferring energy to the less massive component (e.g., Mochnacki, 1981; Yakut \& Eggleton 2005).

\begin{figure}
\begin{center}
\includegraphics[angle=0,scale=1.0]{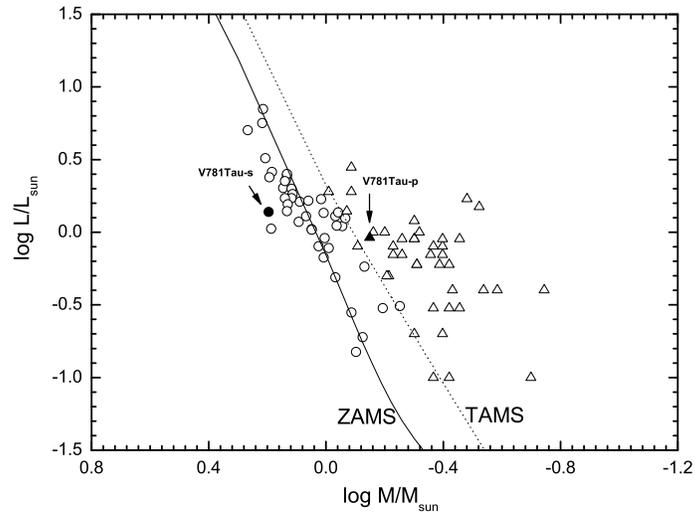}
\caption{Mass-luminosity diagram of V781 Tau. The open triangles and circles represent the primary and secondary components for the W-subtype LTCBs (Yakut \& Eggleton 2005). The solid and dotted lines show the ZAMS and TAMS lines, constructed by the BSE Code (Hurley et al. 2002). }
\end{center}
\end{figure}

According the period changes of 30 W-subtype contact binaries, Qian indicated that systems with decreasing period usually have mass ratios $q<0.4$, while those with increasing period usually have mass ratios $q>0.4$. Then, Qian (2003) expanded his results to both W- and A-subtype contact binaries. Based on the statistical investigation of period changes of contact binaries, Qian (2003) confirmed that the period variation of W UMa-type binaries are correlated with $q$ and $M_1$. He also pointed out that the thermal relaxation oscillation and the variable angular momentum loss can result in W UMa-type systems oscillate around a critical mass ratio.
The mass ratio of V781 Tau ($q_{inverse}=0.45$) is around the critical mass ratio, this binary is a particularly important system and needed more intensive study.

\acknowledgments
This work is partly supported by the National Natural Science Foundation of China (Nos. 11203016, U1431105), and, by the Natural Science Foundation of Shandong Province (No. ZR2014AQ019), and by the Open Research Program of Key Laboratory for the Structure and Evolution of Celestial Objects (No. OP201303). Thank the anonymous referee very much for her/his help, constructive comments and suggestions, which helped to improve this paper.

\section{Compliance with Ethical Standards}
\textbf{Conflict of Interest:} The authors declare that they have no conflict of interest.


\begin{references}

%
\reference{}Applegate, J. H. 1992, ApJ, 385, 621
%
\reference{}Berthold, Th. 1981, IBVS, 1942, 1
%
\reference{}Berthold, Th. 1983, IBVS, 2443, 1
%
\reference{}Cereda, L., Misto, A., Poretti, E., \& Niarchos, P. G. 1988, A\&AS, 76, 255
%
\reference{}Diethelm, R. 1981, BBSAG, 52, 7
%
\reference{}Harris, A. W. 1979, IBVS, 1556, 1
%
\reference{}Hurley, J. R., Tout, C. A., \& Pols, O. R. 2002, MNRAS, 329, 897
%
\reference{}Hu, S. M., Han, S. H., Guo, D. F., \& Du, J. J. 2014, RAA, 14, 719
%
\reference{}Kallrath, J., Milone, E. F., Breinhorst, R. A., Wilson, R. E., Schnell, A., \& Purgathofer, A. 2006, A\&A, 452, 959
%
\reference{}Kreiner, J. M. 2004, AcA, 54, 207
%
\reference{}Lee, J. W., Hinse, T. C., \& Park, J.-H. 2013, AJ, 145, 100
%
\reference{}Li, K., Qian, S.-B., Hu, S.-M., \& He, J.-J. 2014, AJ, 147, 98
%
\reference{}Liu, Q.-Y., \& Yang, Y.-L. 2000, A\&AS, 142, 31
%
\reference{}Lu, W.-X. 1993, AJ, 105, 646
%
\reference{}Lucy, L. B. 1967, Z. Astrophys. 65, 89
%
\reference{}Lucy, L. B. 1967, Z. Astrophys. 65, 89
%
\reference{}Mochnacki, S. W. 1981,ApJ, 245, 650
%
\reference{}Niarchos, P. G. 1983,A\&AS, 53, 13
%
\reference{}O'Connell D. J. K. 1951, MNRAS, 111, 642
%
\reference{}Qian, S. B. 2001, MNRAS, 328, 635
%
\reference{}Qian, S. B. 2003, MNRAS, 342, 1260
%
\reference{}Qian, S.-B., Liu, N.-P., Liao, W.-P., He, J.-J., Liu, L., Zhu, L.-Y., Wang, J.-J., \& Zhao, E.-G., 2013, AJ, 146, 38
%
\reference{}van Hamme, W. 1993, AJ, 106, 2096
%
\reference{}Wilson, R. E., \& Devinney, E. J. 1971, ApJ, 166, 605
%
\reference{}Wilson, R. E. 1990, ApJ, 356, 613
%
\reference{}Wilson, R. E. 1994, PASP, 106, 921
%
\reference{}Yakut, K., Ula\c{s}, B., Kalomeni, B., G\"{u}lmen, \"{O}. 2005, MNRAS, 363, 1272
%
\reference{}Yakut, K., \& Eggleton, P. P. 2005, ApJ, 629, 1055
%
\reference{}Zwitter, T., Munari, U., Marrese, P. M., Pr\v{s}a, A., Milone, E. F., Boschi, F., Tomov, T., \& Siviero, A. 2003, A\&A, 404, 333
%


\end{references}
\end{document}